\documentclass[11pt]{article}
\usepackage{epsfig,latexsym,floatflt,amssymb}

\def\lapprox{\mathrel{\mathop  {\hbox{\lower0.5ex\hbox{$\sim$}
\kern-1.1em\lower-0.7ex\hbox{$<$}}}}}
\def\gapprox{\mathrel{\mathop  {\hbox{\lower0.5ex\hbox{$\sim$}
\kern-1.1em\lower-0.7ex\hbox{$>$}}}}}

\usepackage{axodraw}

\def\bef{\begin{figure}}
\def\eef{\end{figure}}

\newcommand{\be}[1]{\begin{equation}\label{#1}}
\newcommand{\beq}{\begin{equation}}
\newcommand{\ee}{\end{equation}}
\newcommand{\beqn}[1]{\begin{eqnarray}\label{#1}}
\newcommand{\eeqn}{\end{eqnarray}}
\newcommand{\bd}{\begin{displaymath}}
\newcommand{\ed}{\end{displaymath}}

\def\lsim{\raise0.3ex\hbox{$\;<$\kern-0.75em\raise-1.1ex
e\hbox{$\sim\;$}}}
\def\gsim{\raise0.3ex\hbox{$\;>$\kern-0.75em\raise-1.1ex
\hbox{$\sim\;$}}}
\def\simlt{\mathrel{\lower2.5pt\vbox{\lineskip=0pt\baselineskip=0pt
           \hbox{$<$}\hbox{$\sim$}}}}
\def\simgt{\mathrel{\lower2.5pt\vbox{\lineskip=0pt\baselineskip=0pt
           \hbox{$>$}\hbox{$\sim$}}}}
\def\unity{{\hbox{1\kern-.8mm l}}}
\newcommand{\ov}{\overline}
\renewcommand{\to}{\rightarrow}

\newcommand{\mc}{\mathcal}

\newcommand{\gr}{\mathbf}

\renewcommand{\to}{\rightarrow}

\newcommand{\br}{\langle}

\newcommand{\kt}{\rangle}

\newcommand\eV{\mbox{eV}}
\newcommand\keV{\mbox{keV}}
\newcommand\MeV{\mbox{MeV}}
\newcommand\GeV{\mbox{GeV}}

\newcommand\mev{\mbox{\tiny{MeV}}}

\def\lapprox{\mathrel{\mathop  {\hbox{\lower0.5ex\hbox{$\sim$}
\kern-1.1em\lower-0.7ex\hbox{$<$}}}}}
\def\gapprox{\mathrel{\mathop  {\hbox{\lower0.5ex\hbox{$\sim$}
\kern-1.1em\lower-0.7ex\hbox{$>$}}}}}

\def\nn{\nonumber\\}
\def\de{\partial}
\newcommand{\psl}{{p \hspace{-4.4pt} \slash}\;}
\newcommand{\asl}{{a \hspace{-4.4pt} \slash}\;}
\begin{document}

\begin{titlepage}

\begin{flushright}


\end{flushright}
\vspace{1.0cm}

\begin{center}

{\Large \bf Nucleon-Nucleon Bremsstrahlung emission of massive Axions}

\end{center}

\vspace{0.3cm}

\begin{center}

{\large M. Giannotti$^a$ and  F. Nesti$^b$}\\[1em]
\noindent\llap{$^a$}Sezione INFN di Ferrara, I 44100 FE\\
E-mail: {\tt giannotti@fe.infn.it}\\[1ex]
\noindent\llap{$^b$}Dipartimento di Fisica, Universit\`a di L'Aquila,
I-67010 Coppito, AQ, and\\ INFN, Laboratori Nazionali del Gran Sasso,
I-67010 Assergi, AQ, Italy \\ E-mail: {\tt
fabrizio.nesti@aquila.infn.it}\\[2em]
\end{center}

\begin{abstract}

We consider the problem of axion production by bremsstrahlung emission
in a nuclear medium.  The usual assumption of a massless axion is
replaced by more general hypotheses, so that we can describe the
emission process for axions with mass up to a few \MeV.  We point out
that in certain physical situations the contribution from non-zero
mass is non-negligible.  In particular, in the mechanism for the
production of Gamma Ray Bursts via emission of heavy axions the axion
mass ($m_a\sim 1\,\MeV$) is comparable with the temperature of the
nuclear medium, and thus can not be disregarded.
Looking at our results we find, in fact, a fairly considerable reduction of
the axion luminosity in that mechanism.
\end{abstract}

\end{titlepage}

\section{Introduction}
It is largely believed that the Peccei-Quinn (PQ) mechanism~\cite{PQ} must be
realized in nature, since it explains the smallness (or absence) of
the CP-violating term in the strong sector of the SM.  This problem,
known as the Strong-CP problem, is solved dynamically: the CP-violating 
term is driven to zero by the relaxation of a pseudo-scalar
field around its Vacuum Expectation Value (VEV). The field who plays
this role, named axion, is the main prediction (still unverified) of
the theory.  Axions emerge as Pseudo-Goldstone Boson (PGB) modes
associated with the (mostly) spontaneously broken PQ symmetry
$U(1)_{\rm PQ}$.  The PQ or axion decay constant $f_a$, which
corresponds to the energy scale of the spontaneous symmetry breaking,
characterizes almost all the axion properties, on a phenomenological
ground~\cite{kim}.%
\footnote{Strictly speaking, the energy scale of the spontaneous breaking
of the PQ symmetry, say $f_{\rm PQ}$, does not always correspond to the
phenomenological scale $f_a$.  In general $f_a=f_{\rm PQ}/N$, where
$N$ stands for the color anomaly of $U(1)_{\rm PQ}$ current, and the
PQ charges are normalized so that each of the standard fermion
families contributes as $N=1$.  Therefore, in the Weinberg-Wilczek
(WW) model~\cite{WW}, we have $N=N_g$, where $N_g(=3)$ is the number of
fermion families.  The same holds in the
Dine-Fischler-Srednicki-Zhitnitskii (DFSZ) model~\cite{DFSZ}.  Other
models of the invisible axion, e.g.\ the hadronic axion~\cite{KSVZ} or
archion~\cite{archion}, generally contain some exotic fermions and so
$N\neq N_g$.}
More specifically the axion mass is given by the relation
\begin{equation}\label{mass}
	m_a/\eV \simeq 0.62 \frac{10^7\,\GeV}{f_{a}}~,
\end{equation}
while its interactions with fermions are measured by $g_{i}\sim m_i/f_a$,
where $m_i$ represents the fermion mass (e.g.\ $m_e,m_N,...$ for
electrons, nucleons, etc.).

Since the PQ mechanism does not fix $f_a$, the axion phenomenology is
largely model-dependent.  However the presently allowed range for
$f_a$ is rather narrow~\cite{kim}: terrestrial experiments and astrophysical
considerations have in fact excluded all the value of $f_a$ up to
$10^{10}\,\GeV$,%
\footnote{
In the case of the hadronic axion \cite{KSVZ}, 
a small window around $f_a \sim 10^6$ GeV can be also permitted.}
while cosmological considerations about axion non-thermal production
demand the upper bound $f_a \lapprox 10^{11}-10^{12}\,\GeV$.%
\footnote{A possibility of relaxing the cosmological limit is discussed in~\cite{IO}.}

The most stringent lower limits on the axion scale $f_a$ derive from
astrophysics. Indeed, terrestrial experiments exclude values of $f_a$
only up to a few $10^4\,\GeV$.%
\footnote{The somewhat stronger limit, $f_a\simeq 10^5\,\GeV$, emerges
for an axion heavier than two electrons, from the reactor search of
$a\to e^+~e^-$ decay.}
However, this value of the axion constant demands the upper limit on
the axion mass $m_a\lapprox 1\,\keV$, by virtue of relation
(\ref{mass}).  Such a light particle should be emitted from stars of
all varieties, and should thereby affect stellar evolution. Thus the axion
interaction with the stellar matter must be reduced. This
explains the strong limit $\sim 10^{10}\,\GeV$ on the PQ constant,
which is then a direct consequence of relation (\ref{mass}).
However, this relation is not a prediction of the PQ mechanism, and so
does not necessarily apply to axions.  Actually, the only requirement for
the axion field from the PQ mechanism is to dynamically cancel the
CP-violating part of the QCD Lagrangian, and this is still possible
without satisfying relation (\ref{mass}). This would considerably enlarge
the parameter space for the axion, as discussed in~\cite{BGG,GGN}.  It
is then plausible that, in the future axion models, relation~(\ref{mass})
could be re-considered.

We can, in fact, refer to a specific example: in~\cite{GGN}, it is
considered an axion with mass $m_a\sim 1\,\MeV$ and PQ constant 
$f_a\sim 10^{6}\,\GeV$, and it is shown that it can still drive the QCD Lagrangian to its
CP-conserving minimum.  A particle like that can not be excluded by
any phenomenological consideration. In particular, for such a massive
axion, the lower limit on the PQ constant is just the terrestrial one
($f_a\sim$ a few $10^4\,\GeV$), since it can not ruin the stellar
evolution process.

The phenomenology of this {\it non-standard} axion is quite
interesting.  As proposed in~\cite{drago} it can be a key ingredient
in explaining the production of Gamma Ray Bursts (GRBs).  It can be
produced during the merging of two compact objects and then, after
its decay, efficiently transfer the gravitational energy of the
collapsing system into an ultra-relativistic $e^+e^-$ plasma, the
fireball, far from the impact place.  Also, since it can be produced
in the hot core of type II SN, it can decay into $e^+\;e^-$ before
reaching the stellar surface and, by doing so, transfer a huge amount
of energy to a distance of about $1000$ Km from the stellar core,
helping the SN explosion (thermal bomb). On the other hand, because
type Ib/c SN are smaller, some axions are able to leave
their surface and then decay into photons, explaining the observed
events of weak GRBs related to type I SN.

The possible existence of heavy pseudo-scalar particles motivates the
effort to reconsider the most interesting astrophysical axion
processes, removing the usual assumption of zero mass.  A particularly
interesting example is the well studied nucleon-nucleon axion
bremsstrahlung process
\begin{equation}\label{brem}
	N~N\to N~N~a~,
\end{equation}
where $N$ represents a nucleon.  This is the most important axion
production mechanism in the hot and dense core of a SN ($T\sim
30-80\,\MeV$, $\rho \sim (6-10)\times 10^{14}~{\rm g~ cm}^{-3})$, 
and has received much attention in the past years, in particular after
the observation of the neutrino signal from SN1987A.  In fact, an
axion overproduction in the SN core would ruin the temporal structure
of this signal, and this analysis sets the most stringent lower bound
on the PQ constant.  After the pioneering work of Iwamoto, several
other papers discussed the subject~\cite{Iw}, but always in the
hypothesis $m_a\sim 0$. This is clearly very well justified for a
standard axion in the SN core, though not for the "axion-GRBs"
mechanism of ref.~\cite{drago}, where heavy axions are produced in a
medium at a temperature comparable with their mass.

In this paper we discuss the axion emission from nucleon-nucleon axion
bremsstrahlung process, replacing the usual assumption of a massless
axion with more general hypotheses, so that we can describe the
emission process for axions with mass up to a few \MeV.  In addition,
we will consider the effect of the pion mass in the propagator of the
nucleon-nucleon interaction.  Thus we extend the previous analysis,
and with these more general assumptions, the above process can be
studied in a considerably larger set of physical situations.  There
are, in fact, physical conditions in which the standard results can
not be used, while our hypotheses are still valid, as for the limit of
very small axion momentum, and the phenomenology of the "axion-GRBs"
mechanism that we have described above; this last will be our main
reference example throughout the paper. We will revisit it, showing
that the use of the standard results led to an overestimation of the
axion luminosity, for fixed axion-nucleon coupling, by a factor
$3-10$.  This result, however, does not spoil the general idea of
reference~\cite{drago}.

\medskip

The paper is organized as follows: in section 2 we briefly review the
"axion-GRB" mechanism of reference~\cite{drago,GGN}, while in section
3 we describe the emission process in the nucleon-nucleon axion
bremsstrahlung for non-negligible axion mass, and give some numerical
results; finally, in section 4, we summarize the results and add some
comments.  Technical points and some generalizations are discussed in
the appendix.


\section{Axion emission and gamma ray bursts.}
In this section, we briefly review the mechanism of
reference~\cite{drago} that considers a heavy, {\it non-standard}
axion as the key ingredient in the production of Gamma-Ray Bursts
(GRBs).  Since this will be our main example, we will frequently refer
to it throughout the paper as the "axion-GRBs" mechanism.

The most striking feature of GRBs is that an enormous energy, up to
$10^{53-54}$ erg, is released in a few seconds, in terms of photons
with typical energies of several hundred \keV.  The time-structure of
the prompt emission and the afterglow observations well agree with the
fireball model~\cite{meszaros} in which the GRBs originate from the
$e^+ e^-$ plasma that expands at ultrarelativistic velocities,
undergoing internal and external shocks. The Lorentz factor of the
plasma needs to be very large, $\Gamma\sim 10^2$, and this requires a
very efficient acceleration mechanism. In particular, the fireball has
to be formed in a region of low baryonic density so that the $e^+
e^-$ plasma is not contaminated by more massive matter (baryons).
Thus the problem remains how to transform efficiently enough the
available energy into the powerful GRBs.  Due to the low efficiency of
the $\nu\bar\nu \to e^+ e^-$ reaction, the models invoking it as a
source for the GRBs (see, e.g.,~\cite{9_11}) have a lot of difficulty
in reaching such large photon luminosities.

In ref.~\cite{drago}, a more efficient mechanism was proposed that
invokes the $a \to e^+ e^-$ decay, rather than the reaction
$\nu\bar\nu \to e^+ e^-$, where $a$ is a heavy ($m_a\sim\,\MeV$)
pseudoscalar particle.  This can be effectively produced inside the
accretion disks that form after the merging of two compact objects,
like two Neutron Stars~(NS) or NS and Black-Hole~(BH), and,
decaying far from the disk, can efficiently transfer the gravitational
energy of the collapsing system into the ultrarelativistic $e^+e^-$
plasma.  The advantages of this mechanism, with respect to the
$\nu\bar\nu\to e^+e^-$ annihilation, are obvious: first of all, it is
100 percent efficient, since the decaying axions deposit their energy
and momentum entirely in the $e^+e^-$ plasma; in addition, the decay
can take place in the baryon free regions, at distances of 1000 km or
larger, and so the plasma can reach a Lorentz factor $\Gamma\sim 10^2$.

The parameter range needed for this scenario points to an axion-like
particle, with a mass $m_a$ of a few \MeV, coupling to nucleons $g_N\sim
10^{-6}$ and to electrons $g_e\sim 10^{-9}$.  This range of coupling
constants coincides to that of the invisible axion, with the PQ
symmetry breaking scale $f_{a} \sim 10^6\,\GeV$.  However, for such a
value of $f_{a}$, a {\it standard} axion would have a mass of a few \eV,
whereas the needed particle must have a mass of a few \MeV, i.e.\ about a
million times larger.  Hence, its mass must not be constrained by the
standard relation~(\ref{mass}).  However, this ultramassive axion
cannot be excluded by the existing experimental data and astrophysical
limits. Moreover, the relevant parameter window is not far from the
present experimental possibilities, and it can be tested with the reactor and
beam dump experiments in the close future.

A possible candidate for this {\it new} particle could be the "failed"
standard axion, which reaches order \MeV mass via the Planck scale
effects. In this case, however, it cannot be considered for solving the
strong CP-problem.  A more interesting possibility was presented
in~\cite{GGN}, in which the mass relation (\ref{mass}) is changed by
virtue of the axion interaction with a hidden (mirror) sector of
particles.  In this last case, the resulting particle is still an
axion, meaning it still solves the strong-CP problem. 
\medskip

In the hot medium with temperature $T\sim$ a few \MeV and density
$\rho\sim 10^{10}-10^{12} g ~\rm cm^{-3}$, typical of the central zone
of the accretion disk, the nucleons are non-degenerate and
non-relativistic $E_i\sim m+\gr p_i^2/2m$. In these conditions, the
main emission process is the nucleon-nucleon axion bremsstrahlung
$N~N\to N~N~a$.  Even though this process has been extensively studied
in the past, it was always related to the emission of standard axions
from the SN core, where $m_a/T<10^{-10}$. Therefore, the hypothesis
$m_a=0$ was always assumed.  Also, as we will show in the next
section, at the high temperatures inside the SN core, the pion mass in
the propagator of the nucleon-nucleon interaction can be neglected.
On the other hand, this approximation is not justified at the
temperature of a few \MeV.  A careful analysis of the nucleon-axion
bremsstrahlung process is given in the next section. We will show that
the use of the standard results for the axion emission rate
in~\cite{GGN,drago} led to an overestimation of the axion luminosity
for fixed axion-nucleon coupling $g_n$, by a factor $\sim 3-10$.  This
result, however, does not spoil the general idea of ref.~\cite{drago},
since the resulting luminosity is still enough for the production of
the GRBs.

Observe, in addition, that the maximal luminosity obtainable in the
this mechanism remains essentially unchanged, even if the axion and
pion mass effects are taken into consideration.  We can briefly
explain this result, that will be extensively described at the end of
the next section: if the axions are not trapped in the disk, the
luminosity function increases with $g_n$, until the axions start to
interact too strongly with the nuclear matter, and their mean free
path becomes smaller than the accretion disk size. Therefore the
maximal axion luminosity corresponds to a certain value of the
axion-nucleon coupling, $g_{n}^{tr}$, while for a larger coupling the
axions become trapped in the disk, and their emission rate decreases.
If the non-zero mass effects lower the axion luminosity for fixed
coupling $g_n$, they also increase the value of $g_{n}^{tr}$, and
these two effects balance in the resulting maximal luminosity.

\section{Bremsstrahlung emission of heavy axions}

In this section, we study the nucleon-nucleon axion bremsstrahlung
process (\ref{brem}) in the one-pion-exchange approximation (OPE),%
\footnote{The validity of OPE is discussed in ref.~\cite{Turn4}.}
in which nucleons interact with each other by exchanging one pion.  In
addition, we will consider non-degenerate nucleons, and will be
focused, for simplicity, on the $n~n$ bremsstrahlung ($n=$ neutron),
leaving the more general results for the appendix. Wherever it is
possible, throughout this section, we will follow the conventions used
in~\cite{Turn3} and~\cite{RS}. In particular, we will use "$p_i$" and
"$a$" respectively for the nucleons and for the axion four-momentum,
and "$\omega_a$" for the axion energy.

\medskip

All the observables we are interested in can be expressed in terms of
the differential axion emission rate:%
\footnote{A note on terminology: throughout this paper, the term
"axion number emission rate", or simply "axion emission rate" $\mc N$,
indicates the number of axions emitted in the process per unit time
and per unit volume, while the "axion energy emission rate" $Q$ refers
to the energy emitted per unit time and volume (axion luminosity per
unit volume).  The latter is also called "axion energy loss rate"
in~\cite{RS,raffBook}, "axion emission rate" in~\cite{Turn3}, and
"axion volume emission rate" in~\cite{Turn1}.}
\begin{equation}\label{dN}
	d\mc N=d\Pi_a \int d\Pi \{\mc{M}^2\} f_1f_2 (2\pi)^4\delta(p_1+p_2-p_3-p_4-a)\,,
\end{equation}
where $d\Pi=\prod d^3 \gr p_i/[(2\pi)^3 2E_i]$ is the
Lorentz-invariant phase-space volume element for the four nucleons,
while $d\Pi_a=d^3 \gr a/[(2\pi)^3 2\omega_a]$ refers to the axion.
The occupation numbers of the nucleons $f_i\equiv f(\gr p_i)$ 
are given by the Maxwell-Boltzmann distribution:
\begin{equation}\label{distr}
	f(\gr p)=\frac{N_B}{2}\left(\frac{2\pi}{m T}\right)^{3/2}e^{-\gr p^2/2mT},
\end{equation}
with normalization $2\int f(\gr p)d^3 \gr p/(2\pi)^3=N_B$.  The Pauli
blocking factors $(1-f_3)(1-f_4)$ have been omitted, since we are
considering non-degenerate nucleons.  From the above definition, the
axion number emission rate $\mc N$ and the axion energy emission rate
$Q$ are respectively:
\begin{equation}\label{NQ}
	\mc N=\int d\mc N,\qquad Q=\int\omega_a d \mc N\,,
\end{equation}
so that $Q/\mc N$ gives the mean energy of the emitted axions.

For massless axions, the matrix elements squared summed over spins is:
\begin{equation}\label{matrix0}
	\{\mc{M}_0^2\}\equiv
	S\sum_{\rm spin}|\mc{M}_0|^2=\frac{64\pi^2\alpha_\pi^2}{3 m^2}~g_{n}^2
	\left(\frac{|\gr{k}|^4}{(|\gr{k}|^2+m_\pi^2)^2}+\frac{|\gr{l}|^4}{(|\gr{l}|^2+m_\pi^2)^2}+
	\frac{|\gr{k}|^2|\gr{l}|^2-3|\gr{k}\cdot\gr{l}|^2}{(|\gr{k}|^2+m_\pi^2)(|\gr{l}|^2+m_\pi^2)}\right),
\end{equation}
where $g_n\simeq m/f_a$, $\alpha_\pi=(2fm/m_\pi)^2/4\pi\sim 15$
($f\sim 1$ is a phenomenological constant that accounts for the
nucleon-pion interaction) and $S$ is the usual symmetry factor:
$S=1/n!$ for $n$ identical particles in the final state.%
\footnote{The above result, (\ref{matrix0}), is the same as the result
in~\cite{Turn3} and ~\cite{RS}. Observe, though, that in~\cite{RS}
$C_{an}$ is used in place of $g_n$, with $g_n=(2m/f_a)C_{an}$.  In
addition, it is defined $\alpha_a$ such that $g_n^2=4\pi\alpha_a
C_{an}^2$.}
The three-momentum transfers in the "direct" and "exchange" diagrams
are indicated respectively with $\gr k =\gr p_2-\gr p_4$ and $\gr l
=\gr p_2-\gr p_3$ (see figure~\ref{fig:diagrams} in the appendix).
Observe that $\gr k^2\sim 3 m T$ so that $\gr k^2/m_\pi^2\sim0.15~
(T/1\,\MeV)$.  Certainly, in a medium with $T\sim$ a few $10\,\MeV$,
like the SN core, the pion mass can be neglected,%
\footnote{We will frequently refer to the "massless pion limit" or
"negligible pion mass effects", etc.  throughout the paper. With these
expressions we will always intend $\gr k^2/m_\pi^2\ll 1$.}
and the expression for the matrix element squared can be considerably
simplified:
\begin{equation}\label{matrPi0}
	\{\mc{M}_0^2\}_{m_\pi\to 0}~=\frac{64\pi^2\alpha_\pi^2}{m^2}~g_{n}^2
	(1-\hat{|\gr{k}}\cdot\hat{\gr{l}}|^2)\,,
\end{equation}
where $\hat{\gr{k}}=\gr{k}/|\gr{k}|$ and $\hat{\gr{l}}=\gr{l}/|\gr{l}|$.
In this case, from the definitions in (\ref{NQ}), one gets:
\begin{equation}\label{Q0}
	Q_0=\frac{32(3-\beta)}{105}\frac{\rho^2 T^{7/2}\alpha_\pi^2}{\pi^{3/2}m^{13/2}}~g_n^2
	\simeq 5.75 \times 10^{42} T_{\mev}^{7/2}\;\rho_{12}^2\;g_n^2~{\rm erg~ cm^{-3}~s^{-1}}\,,
\end{equation}
where $\beta=1.31$ accounts for the contribution of $\hat{|\gr{k}}\cdot\hat{\gr{l}}|^2$ 
(see the appendix), while
$T_{\mev}=T/1{\MeV}$ and $\rho_{12}=\rho/(10^{12}{\rm g~cm^{-3}})$.
Analogously,
\begin{equation}\label{N0}
\mc N_0=\frac7{16}\left(\frac{3-\beta^\prime}{3-\beta}\right)\frac{Q_0}{T}=
1.84\times 10^{48}  T_{\mev}^{5/2}\;\rho_{12}^2\;g_n^2~{\rm cm^{-3}~s^{-1}}\,,
\end{equation}
where $\beta^\prime=1.02$ is defined similarly to $\beta$ (see the appendix). Finally
\begin{equation}\label{E0}
\ov \omega_0=Q_0/\mc N_0\simeq 1.95\,T
\end{equation}
is the mean energy of the emitted axions, in the limit of negligible
pion and axion masses.

\medskip

If the assumption that the axion is massless is removed,
eq.~(\ref{matrix0}) can no longer describe the axion bremsstrahlung.
This can be understood by the following argument. If $m_a\neq0$, we
can consider the limit of $|\gr a|\ll\omega_a$.  In this limit, we
expect $\mc M^2\to0$. In fact, as any Goldstone mode, axions interact
only derivatively and thus the axion-nucleon coupling must vanish for
vanishing axion three-momentum~$\gr a$.  Instead, $\mc M^2_0$ does not
depend on $\gr a$, and so it can not describe the correct behavior in
the above limiting situation. The range of validity of the standard
result is indeed $\omega_a\gg m_a$, which means $|\gr a|\sim\omega_a$.

\pagebreak[3]

We have, then, calculated the matrix element squared assuming, in place of
$m_a\sim 0$, the more general hypotheses:

\begin{itemize}
\item[\emph{i})] The nucleon mass is much larger than both the temperature,
$m\gg T$, and the axion mass, $m\gg m_a$;

\item[\emph{ii})] The axion mass is negligible with respect to the
momentum transfer $m_a\ll \sqrt{mT}\sim 30 \,\MeV\,(T_{\mev})^{1/2}$.
\end{itemize}
If $T$ is less than a few $10\,\MeV$, the above hypotheses are easily
satisfied in the range $m_a\sim 0$, up to $m_a\sim$ a few $T$.  Note
also that, since the kinetic energy of the emitted axions is $\sim 2-3
T$, hypotheses (\emph{i}) and (\emph{ii}) imply that the axion
three-momentum is negligible with respect to that of the nucleons.  

Assuming the hypotheses above we have found:
\begin{equation}\label{matrix}
	\{\mc{M}^2\}\equiv 
	S\sum_{\rm spin}|\mc{M}|^2=\left(\frac{\gr a^2}{\omega_a^2}\right)\{\mc{M}_0^2\}\,.
\end{equation}
Therefore, the more general assumptions above have led to a very
simple modification of expression~(\ref{matrix0}).  The correction
factor $\gr a^2/\omega_a^2=(1-m_a^2/\omega_a^2)$, which actually
is the velocity squared of the emitted axion, becomes fairly
irrelevant ($\sim1$) in the limit $m_a\ll \omega_a$.  Observe that the
result (\ref{matrix}) confirms the expected behavior for small axion
momentum.

\medskip

\begin{figure}[t]%
\noindent\includegraphics[width=7.5cm,angle=0]{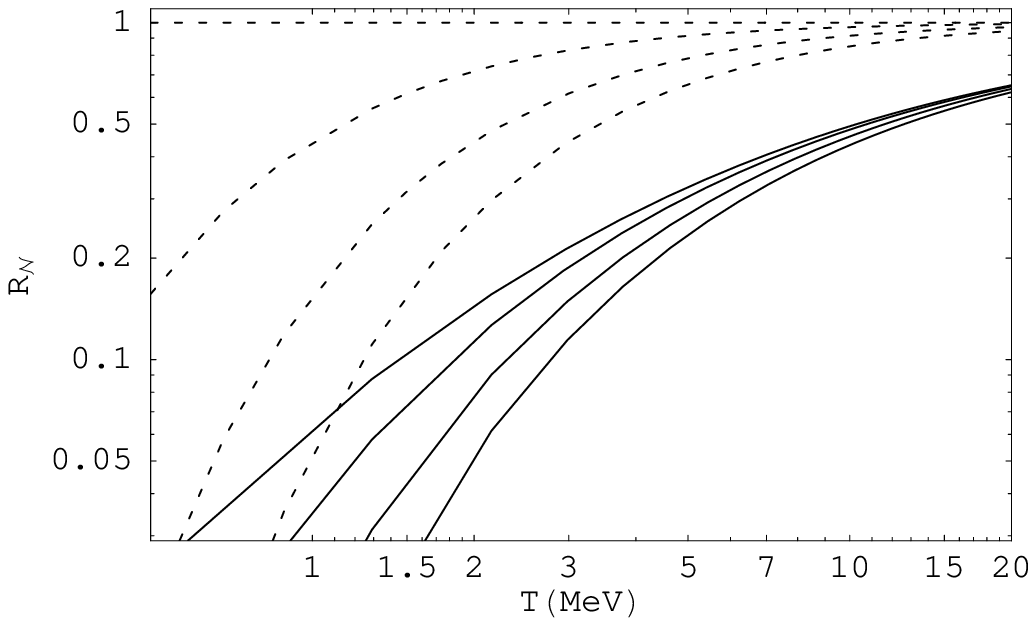}%
\noindent\includegraphics[width=7.5cm,angle=0]{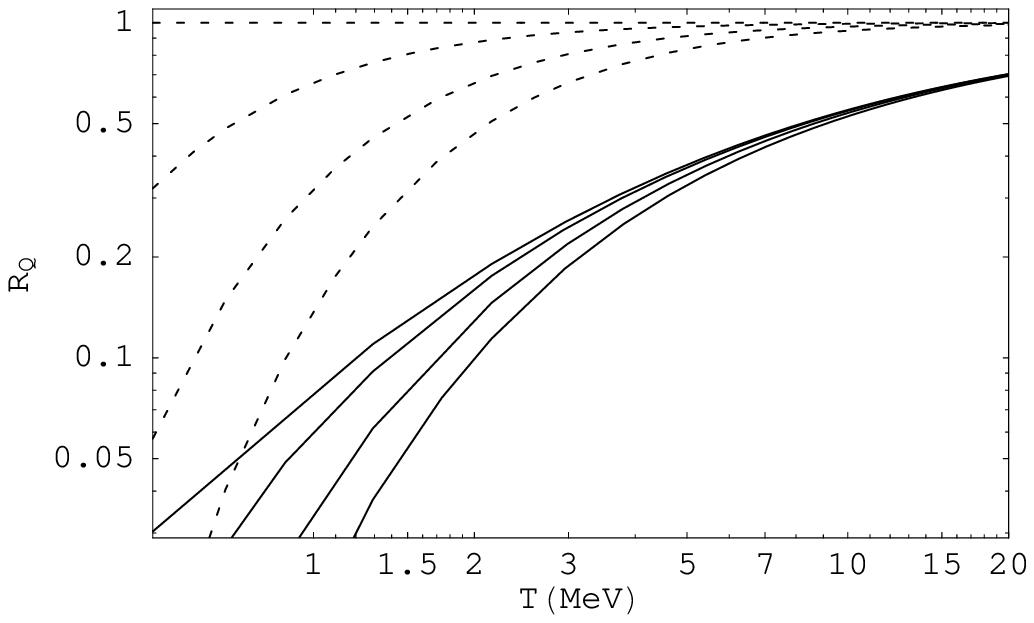}%
\caption{\protect\small Effects of non-zero axion and pion masses for
the axion number emission rate (left panel) and energy emission rate
(right panel).  Continuous lines represent $R_{\mc N}(m_a,m_\pi)=\mc
N(m_a,m_\pi)/\mc N(0,0)$ (left), and
$R_{Q}(m_a,m_\pi)=Q(m_a,m_\pi)/Q(0,0)$ (right), for different values
of the axion mass: $m_a=0,1,2,3\,\MeV$ from the top to the bottom
lines. Dashed lines represent, for the same values of~$m_a$, the
effect of the axion mass alone (i.e.\ the pion mass is set to zero).
\label{fig:R}}%
\end{figure}

Using the complete matrix element squared (\ref{matrix}) we have
calculated, numerically, the axion number and energy rates (\ref{NQ}).
Our results are shown in figure~\ref{fig:R}, where we plot the effects
of pion and axion mass on the emission and the energy emission rates.
In the left panel, we show the correction to the axion emission rate
$$R_{\mc N}(m_a,m_\pi)=\mc N(m_a,m_\pi)/\mc N(0,0)\,,$$ 
where $\mc N(m_a,m_\pi)$ is defined in (\ref{NQ}) 
(see also relation (\ref{N}) in the appendix), while $\mc N(0,0)$ is given in
(\ref{N0}). We plot this reduction factor for the value of the
pion mass $m_\pi=135$ \MeV and for different values of the axion mass, $m_a=0$, $1$, $2$,
$3$ \MeV (solid lines).  As a comparison, we also show 
$R_{\mc N}(0,m_a)$, $m_a=0$, $1$, $2$, $3$ \MeV (dashed lines), where the pion mass is neglected.

Similarly, in the right panel of figure~\ref{fig:R} we consider the
 axion energy emission rate
$$R_{Q}(m_a,m_\pi)=Q(m_a,m_\pi)/Q(0,0)\,,$$ 
for $m_a=0$, $1$, $2$, $3$ \MeV, with (solid lines) and without
(dashed lines) the pion mass contribution.  Observe that both $R_{\mc
N}$ and $R_Q$ depend on the ratios $m_a/T$ and $m_\pi^2/mT$, and not
separately on $m_a$, $m_\pi$ and $T$.

As anticipated, in the temperature range of interest for the
mechanism~\cite{drago}, $T=$2--$5\,\MeV$, the effect of the pion mass
is important, and can be up to one order of magnitude, as expected
from the pion propagator suppression. In fact, for $T\lapprox5\,\MeV$,
$m_\pi>|\gr k|$ and $|\gr k|^4/(|\gr k|^2+m_\pi^2)^2$ can be roughly
approximated as $|\gr k|^4/m_\pi^4\sim0.1-0.6$, for $T=2-5\,\MeV$.%
\footnote{This effect was estimated approximately in
ref.~\cite{Turn1}, and it is also discussed in ref.~\cite{RS}, both
times in the massless axion approximation.}
Besides this, also the axion mass induces a suppression, which is
about a factor 2, for $m_a\sim T\sim$ a few \MeV.  Clearly, when the
mass is greater than the temperature, the suppression becomes
exponential because of the nucleons Boltzmann distribution.

\bigskip

The ratio $Q/\mc N$ gives the mean energy $\ov \omega_a$ of the
emitted axions.  In figure~\ref{fig:kin} we present the average
kinetic energy divided by the temperature: $\ov E_{\rm
kin}/T=(\ov\omega_a-m_a)/T$. This can be compared with the case of
thermal axions, for which $\ov E_{\rm kin}/T$ is either $3/2$
(non-relativistic) or $3$ (relativistic).  The dashed line is
calculated, again, neglecting the pion mass. Thus it corresponds to
the limit of high temperature.%
\footnote{Moreover, since $T$ is high, it describes the situation with
$m_a$ very large ($m_a$ up to 5 $T$).}  In this limit, and for
massless axions,  the dashed line of
figure~\ref{fig:kin} reproduces the result (\ref{E0}) as $\ov E_{\rm
kin}/T\simeq 1.95$ (left endpoint).  We see that, in this case, the emitted axions are
less energetic than thermal axions.  At lower temperatures, the pion
contribution is important, and the mean kinetic energy per temperature
increases.  In fact, for massive pions, low energy processes are more
difficult, and the axions are emitted only by the most energetic
nucleons.  For example, at $T=1\,\MeV$ and for massless axions, $\ov
E_{\rm kin}\simeq 2.48\; T$, about $30\%$ more than in the high
temperature limit.  The dependence of $\ov E_{\rm kin}/T$ on the axion
mass is less strong.  It changes not more that $5\%$ for the $m_a$ in
the range $0-5T$.



\makeatletter
\newcommand{\DOUBLEFIGURE}[5][ht]{\@dblfloat{figure}[#1]\centerline{%
                \parbox{.48\textwidth}{\centerline{\epsfig{file=#2}}}~~~
                \parbox{.48\textwidth}{\centerline{\epsfig{file=#3}}}}
                \centerline{\parbox[t]{.48\textwidth}{\caption{#4}}~~~
                \parbox[t]{.48\textwidth}{\caption{#5}}}\end@dblfloat}
\makeatother

\DOUBLEFIGURE[t]{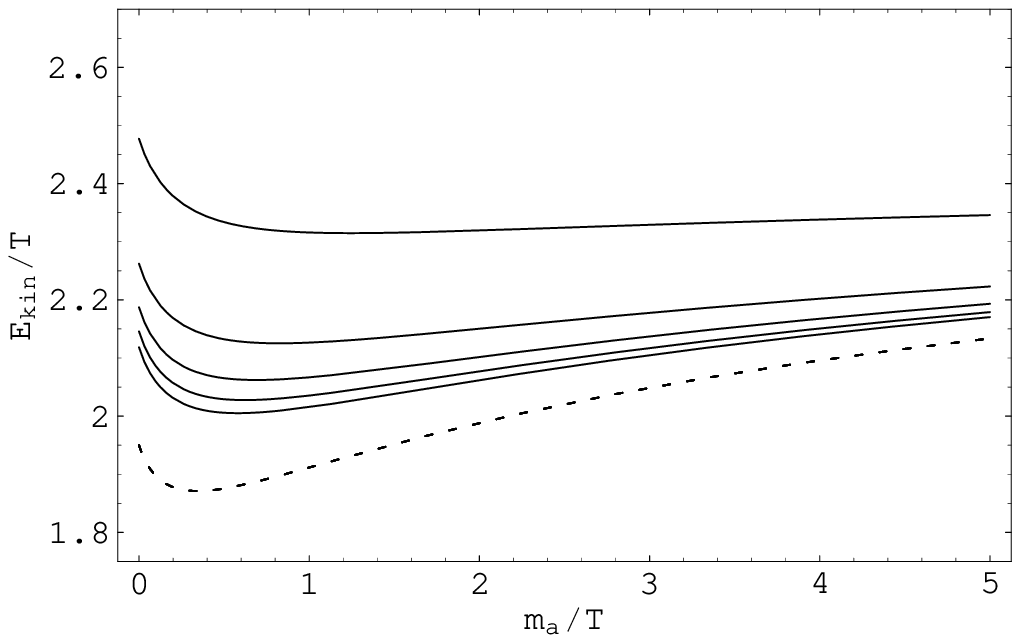,height=4.3cm}{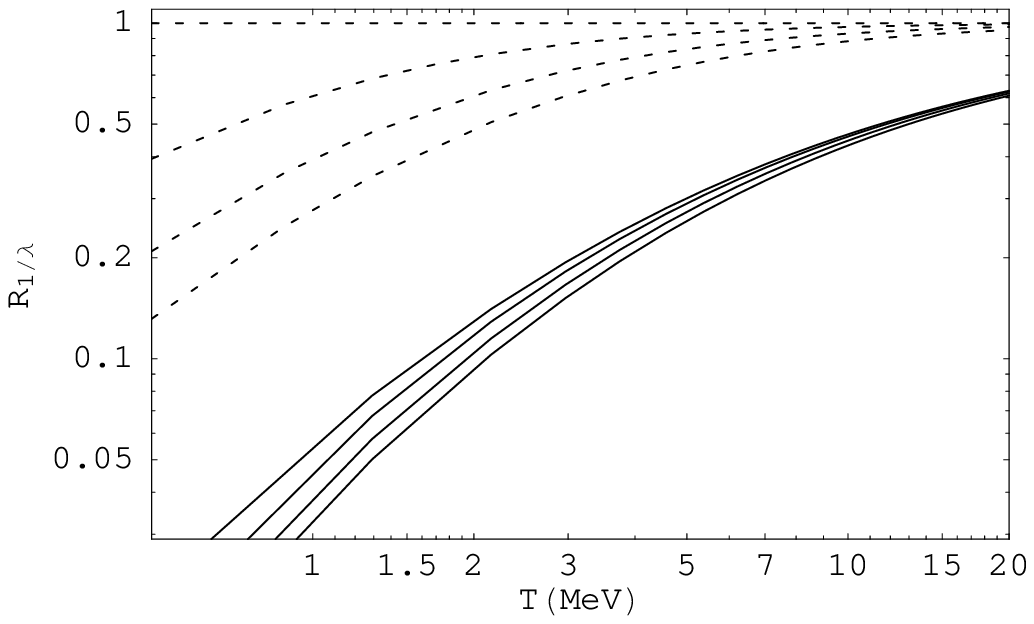,height=4.57cm}
{%
\protect\small
Mean kinetic energy of the emitted axions, divided by the temperature,
with (continuous) or without (dashed) the effect of nonzero pion mass,
for different values of the temperature: from the top to the bottom
solid line $T=1$, 5, 9, 13, 17$\,\MeV$.  Note that the dashed line,
where $m_\pi=0$, also represents the very high temperature limit
$3mT\gg m_\pi^2$.\label{fig:kin}%
}
{%
\protect\small 
Reduction of the inverse mean free path of the {\em emitted} axions,
due to non-zero axion and pion masses.  Continuous lines represent
$R_{1/\lambda}(m_a,m_\pi)=\lambda^{-1}(m_a,m_\pi)/\lambda^{-1}(0,0)$
for different values of the axion mass: from the top to the bottom
line $m_a=0,1,2,3\,\MeV$.  Dashed lines represent
$R_{1/\lambda}(m_a,0)$ for the same values of the axion mass.
\label{fig:RK}
}%


\bigskip

Let us finally discuss the axion mean free path $\lambda$, in the
nuclear medium. This is defined~as
\begin{equation}\label{k}
\lambda^{-1}=\frac1{2|\gr a|}\frac{d\mc N(-a)}{d\Pi_a}=
\frac1{2|\gr a|}\int\!d\Pi\,\{\mc M^2\}f_1f_2(2\pi)^4\delta(p_1+p_2-p_3-p_4+a)\,,
\end{equation}
that reduces to the definition (2) in \cite{Turn1} in the limit of
zero axion mass.  The notation $d\mc N(-a)$ means that the axion
four-momentum must be taken with the opposite sign in the $\delta$
function, with respect to expression (\ref{dN}). In fact, in this
case, the relevant process is the axion absorption $N~N~a\to N~N$,
with the axion in the initial state.

Of course $\lambda$ is a function of the axion energy. 
A simple case, which is particularly interesting for the discussion below, is
when the mean free path is large enough for the absorbed axions to be non-thermal.
In this case, the average axion energy was calculated above (see figure~\ref{fig:kin}).
With this assumption, in the limit of negligible pion and axion mass, 
$\lambda$ can be well-approximated by the relation (see the appendix):
\begin{equation}\label{lambda0}
	 \lambda_0^{-1}=
	 4.2\times 10^6~T_{\mev}^{-1/2}\;\rho_{12}^2\;g_n^2~{\rm cm^{-1}}\,,
\end{equation}
where we used $\ov\omega_a=1.95\;T$.

\pagebreak[3]

In figure~\ref{fig:RK} we have shown the reduction of the inverse mean
free path due to finite axion and pion mass effects:
\begin{equation}\label{RL}
	R_{1/\lambda}(m_a,m_\pi)=\lambda^{-1}(m_a,m_\pi)/\lambda^{-1}(0,0)\,.
\end{equation}
We notice that the pion mass contribution is similar in the
absorption and in the emission processes. Thus, at low temperatures,
we expect a suppression of (\ref{RL}) by a factor $\sim |k|^4/m_\pi^4$.
On the other hand, the axion mass plays a different role in the two processes.
In the appendix it will be shown that non-zero axion mass 
effects can be approximately accounted as
\begin{equation}\label{lma}
	 R_{1/\lambda}(m_a,0)\simeq
	 0.6~(1-m_a^2/\ov\omega_a^2)^{1/2}~e^{\ov\omega_a/2T} K_1(\ov\omega_a/2T)\,,
\end{equation}
where $K_1$ is the modified Bessel function of the second kind. 
For example, for $T=2$ \MeV, $m_a=1$ \MeV, we find $R_{1/\lambda}\simeq0.8$,
in accordance with figure~\ref{fig:RK}.
The contribution from the pion mass lowers the above result to 
$\sim 0.1$ (using $\sim |k|^4/m_\pi^4\sim0.1$ for $T=2$ \MeV, we would predict $R_{1/\lambda}\sim0.08$).

\begin{figure}[t]%
\centerline{\includegraphics[width=9.5cm,angle=0]{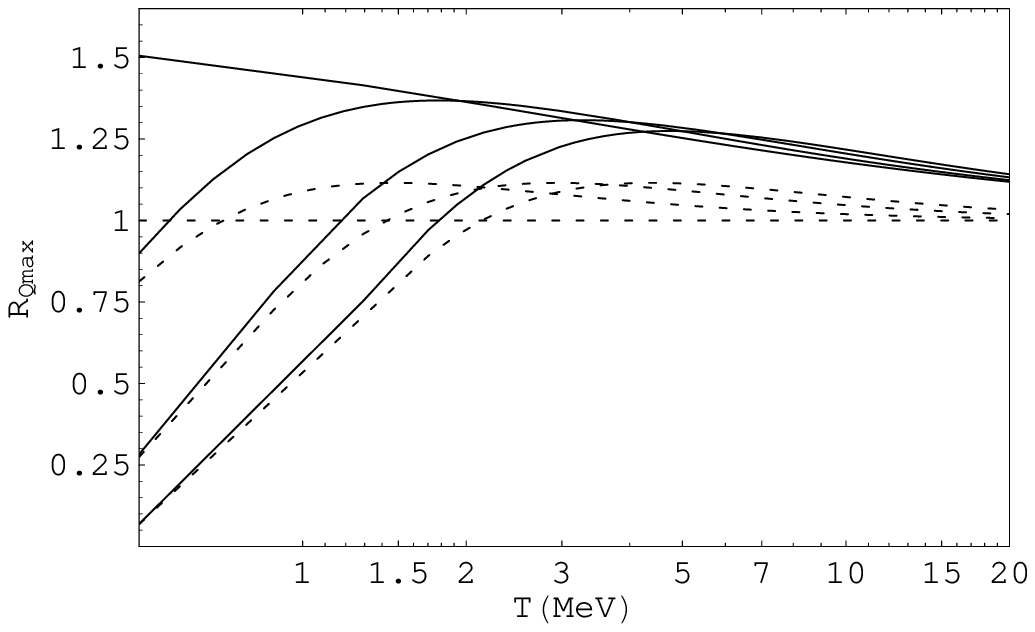}}%
\caption{\protect\small Modification of the maximal GRB luminosity
from non-zero axion and pion mass effects: $R_{Q{\rm MAX}}=Q_{\rm
MAX}(m_a,m_\pi)/Q_{\rm MAX}(0,0)$.  Dashed lines are calculated
omitting the effect of nonzero pion mass. The values of the axion mass
are $m_a=0$, 1, 2, $3\,\MeV$.
\label{fig:RQMAX}}%
\end{figure}

We can now understand better what was discussed at the end of the last section about
the "maximal luminosity" obtainable in the "axion-GRBs" mechanism of ref.~\cite{drago}.
As explained, in order to have a large luminosity,
the axions must be not trapped in the accretion disk.  This is the case
for $g_n<g_n^{tr}$, so that the axion mean free path does not exceed
the accretion-disk radius. 
In the limit of negligible pion and axion mass, 
from (\ref{lambda0}) we get the relation for $g_n^{tr}$
in terms of the accretion disk ratio $R_{100}=R/100 {\rm Km}$, reported in~\cite{GGN,drago}:
$g_n^{tr}\simeq1.5\times 10^{-7}R_{100}^{-1/2}\rho_{12}^{-1}T_{\mev}^{1/4}$.%
\footnote{Observe that in references~\cite{GGN,drago} this result was slightly overestimated
because of the assumption $\ov E_{\rm kin}\simeq 3\;T$.}
The maximal axion luminosity (per volume) corresponds, roughly, to
$Q_{\rm Max}\equiv Q(g_n=g_n^{tr})$. We have shown that both the axion
and pion mass contribute to reducing the axion luminosity, for fixed
axion-nucleon coupling.  However, these effects reflect also on the
axion mean free path and consequently on $g_n^{tr}$, which increases,
balancing the reduction of the axion energy emission rate.  This is
clear comparing figure~\ref{fig:RK} with
figure~\ref{fig:R}-right. Suppose, for example, that $T=2$ \MeV and
$m_a=1$ \MeV.  We find a reduction of the axion luminosity by a factor
of $7$, compensated by an increment of the axion mean free path by a
factor of about $9$. Thus we find that $Q_{\rm Max}$ is about $1.25$
times larger, in the above conditions. This is confirmed in
figure~\ref{fig:RQMAX}, where we have presented the effect of the
axion and pion masses on the maximal axion luminosity, as a function
of the temperature.  As anticipated, there are no significant changes
in the resulting maximal luminosity, except for a factor of $\sim 2-4$
in the region of very low temperatures.
 
\section{Conclusions}

The existence of a heavy {\it non-standard} axion with mass
$\sim\,\MeV$ can not be excluded by any phenomenological
consideration. Indeed, explicit models have been considered in the
past which relax the relation (\ref{mass}) between $f_a$ and $m_a$ and
which allow the axion mass to be very large.  Interestingly, a quite
efficient mechanism for the GRB production was proposed
in~\cite{drago}, which requires a massive axion-like particle to
transfer gravitational energy into the $e^+ e^-$ fireball.  The
conditions in which this particle is produced, temperature $T\sim$ a
few \MeV, and density $\rho\sim 10^{10}-10^{12} {\rm g\;cm^{-3}}$,
favor the nucleon-nucleon axion bremsstrahlung production.  Even if
this process has been studied extensively, the problem of the emission
of an axion with mass not negligible with respect to the temperature
of the medium has never been considered in the past. Thus, in
particular, the luminosity reported in~\cite{drago} is overestimated
by a factor of $3-10$.

In this paper, we have re-studied the problem of nucleon-nucleon axion
bremsstrahlung, relaxing some of the old hypotheses in order to
enlarge the possibility for applications.  In this section, we
summarize the most interesting points of this paper:

\emph{i}) The usual assumption of negligible axion mass has been
replaced with more general hypotheses: \emph{a}) the nucleon mass is much
greater than both the temperature, $m\gg T$, and the axion mass, $m\gg
m_a$; \emph{b}) the axion mass is negligible with respect to the
momentum transfer $m_a\ll \sqrt{mT}\sim 30 (T_{\mev})^{1/2}\,\MeV$.
If $T$ is less than a few $10\,\MeV$, the above hypotheses are easily
satisfied in the range $m_a\sim 0$, up to $m_a\sim T$ or so.  As we
have shown, (\emph{a}) and (\emph{b}) imply that the axion
three-momentum is negligible with respect to that of the nucleons.

\emph{ii}) We have computed the axion number and energy emission rate
$\mc N$ and $Q$ in the conditions of point (\emph{i}).  Besides the
obvious \emph{kinematic} suppression due to the reduction of the axion
phase space, there is also a less trivial \emph{dynamical} effect of
non-zero axion mass, due to the change (\ref{matrix}) of the matrix
element squared.  We also have considered the effects of finite pion
mass, which are important for temperatures below $\sim 6\,\MeV$.  The
final results are presented in figure~\ref{fig:R}, in which we show
the axion number emission rate (left panel) and energy emission rate
(right panel), normalized to the standard one ($m_a=m_\pi=0$), both
considering (continuous lines) or neglecting (dashed lines) the
effects of nonzero pion mass.  As we see, in the temperature range of
interest for the mechanism~\cite{drago}, 2--$5\,\MeV$, the suppression
due to $m_\pi\neq 0$ is important, and amounts to a factor from $\sim
2$ up to one order of magnitude, as expected from the pion propagator
suppression.  On the other hand, the axion mass induces a suppression
which is around a factor of 2, for an axion mass of a few \MeV, and
becomes much larger when the mass exceeds the temperature, mainly
because of the Boltzmann exponential suppression.

\emph{iii}) The two effects of non-negligible axion and pion mass show
that the result of reference~\cite{drago} overestimates the luminosity
for fixed axion-nucleon coupling by roughly an order of magnitude.

\emph{iv}) The reduction of the nucleon-axion interaction rate induces
an increment of the axion mean free path in the medium.  Our numerical
results are presented in figure~\ref{fig:RK}, where we show the
behavior of
$R_{1/\lambda}(m_a,m_\pi)=\lambda^{-1}(m_a,m_\pi)/\lambda^{-1}(0,0)$,
with $\lambda(m_a,m_\pi)$ the mean free path for fixed axion and pion
mass.  Again, for the dashed lines the contribution of finite pion
mass was neglected.

\emph{v}) Even if the actual luminosity for fixed axion-nucleon
coupling is reduced by almost an order of magnitude, the maximal
luminosity obtainable in the "axion-GRBs" mechanism of
ref.~\cite{drago} does not considerably change.  In fact, the
reduction of the axion emission rate is compensated by the increasing
of the mean-free path which ultimately allows a larger value for the
axion-nucleon coupling constant, without trapping the axions in the
disk. We note however that for this to happen one should be able to
take a larger axion-nucleon coupling, which is usually not easy, since
$g_n\sim m/f_a$ is only slightly model dependent.

\emph{vi}) We notice, to conclude, that our corrections to the results
of ref.~\cite{GGN,drago}, even if non-negligible, do not spoil the
general idea of the "axion-GRBs" mechanism.  As remarked in
ref.~\cite{drago}, the emitted axions can still produce the GRBs, even
if their luminosity is reduced by one order of magnitude.

We also mention that for an axion such as the one described
in~\cite{drago} and~\cite{BGG,GGN}, the brems
strahlung process inside
the SN core is very well described by the standard formula~(\ref{Q0}).
In particular, the effect of axion and pion masses on the limit on the
axion-nucleon coupling, reported in ref.~\cite{BGG,GGN}, is
negligible.

\section*{Acknowledgments}

We are grateful to Z. Berezhiani and Denis Comelli for very useful
discussions and suggestions.  We also thank A. Drago for many
explanations about the mechanism of GRBs production via emission of
axion-like particles.  We are finally thankful to Elizabeth Price for
the careful reading of this paper.

\bigskip

\section*{Appendix}

Here we discuss in detail some technical points necessary for the
analysis of the nucleon-nucleon axion bremsstrahlung process.

First of all, we consider closely the matrix element squared for the
process $N~N\to N~N~a$, for a massive axion, assuming the One Pion
Exchange approximation (OPE).

\begin{floatingfigure}{20em}%
\label{fig:diagrams}
\SetScale{.57}
\setlength\unitlength{.57\unitlength}
\noindent
\begin{picture}(350,80)(20,-5)
%
\ArrowLine(50,10)(100,10)  \ArrowLine(100,10)(150,10)     	
\ArrowLine(50,50)(100,50)\ArrowLine(100,50)(150,50)		
\DashLine(100,10)(100,50){2}					
\DashLine(115,50)(150,70){2}  					
\Text(40,10)[]{$p_2$} \Text(40,50)[]{$p_1$}
\Text(160,10)[]{$p_4$} \Text(160,50)[]{$p_3$}
\Text(110,20)[]{$\pi$} \Text(90,30)[]{$k$} 
\Text(158,70)[]{$a$}
%
\ArrowLine(240,10)(290,10)  \ArrowLine(290,10)(340,10)     	
\ArrowLine(240,50)(290,50)\ArrowLine(290,50)(340,50)		
\DashLine(290,10)(290,50){2}					
\DashLine(305,50)(340,70){2}  					
\Text(230,10)[]{$p_2$} \Text(230,50)[]{$p_1$}
\Text(350,10)[]{$p_3$} \Text(350,50)[]{$p_4$}
\Text(300,20)[]{$\pi$} \Text(280,30)[]{$l$} 
\Text(348,70)[]{$a$}
\end{picture}%

\vspace*{2ex}

\noindent
\begin{picture}(350,80)(20,-5)
%
\ArrowLine(50,10)(100,10)  \ArrowLine(100,10)(150,10)     	
\ArrowLine(50,50)(100,50)\ArrowLine(100,50)(150,50)		
\DashLine(100,10)(100,50){2}					
\DashLine(115,10)(150,30){2}  					
\Text(40,10)[]{$p_2$} \Text(40,50)[]{$p_1$}
\Text(160,10)[]{$p_4$} \Text(160,50)[]{$p_3$}
\Text(90,30)[]{$k$} 
\Text(158,30)[]{$a$}
%
\ArrowLine(240,10)(290,10)  \ArrowLine(290,10)(340,10)     	
\ArrowLine(240,50)(290,50)\ArrowLine(290,50)(340,50)		
\DashLine(290,10)(290,50){2}					
\DashLine(305,10)(340,30){2}  					
\Text(230,10)[]{$p_2$} \Text(230,50)[]{$p_1$}
\Text(350,10)[]{$p_3$} \Text(350,50)[]{$p_4$}
\Text(280,30)[]{$l$} 
\Text(348,30)[]{$a$}
\end{picture}%

\vspace*{2ex}

\noindent
\begin{picture}(350,80)(20,-5)
%
\ArrowLine(50,10)(100,10)  \ArrowLine(100,10)(150,10)     	
\ArrowLine(50,50)(100,50)\ArrowLine(100,50)(150,50)		
\DashLine(100,10)(100,50){2}					
\DashLine(65,50)(100,70){2}  					
\Text(40,10)[]{$p_2$} \Text(40,50)[]{$p_1$}
\Text(160,10)[]{$p_4$} \Text(160,50)[]{$p_3$}
\Text(90,30)[]{$k$} 
\Text(108,70)[]{$a$}
%
\ArrowLine(240,10)(290,10)  \ArrowLine(290,10)(340,10)     	
\ArrowLine(240,50)(290,50)\ArrowLine(290,50)(340,50)		
\DashLine(290,10)(290,50){2}					
\DashLine(255,50)(290,70){2}  					
\Text(230,10)[]{$p_2$} \Text(230,50)[]{$p_1$}
\Text(350,10)[]{$p_3$} \Text(350,50)[]{$p_4$}
\Text(280,30)[]{$l$} 
\Text(298,70)[]{$a$}
\end{picture}%

\vspace*{2ex}

\noindent
\begin{picture}(350,80)(20,-5)
%
\ArrowLine(50,10)(100,10)  \ArrowLine(100,10)(150,10)     	
\ArrowLine(50,50)(100,50)\ArrowLine(100,50)(150,50)		
\DashLine(100,10)(100,50){2}					
\DashLine(65,10)(100,-10){2}  					
\Text(40,10)[]{$p_2$} \Text(40,50)[]{$p_1$}
\Text(160,10)[]{$p_4$} \Text(160,50)[]{$p_3$}
\Text(90,30)[]{$k$} 
\Text(108,-10)[]{$a$}
%
\ArrowLine(240,10)(290,10)  \ArrowLine(290,10)(340,10)     	
\ArrowLine(240,50)(290,50)\ArrowLine(290,50)(340,50)		
\DashLine(290,10)(290,50){2}					
\DashLine(255,10)(290,-10){2}  					
\Text(230,10)[]{$p_2$} \Text(230,50)[]{$p_1$}
\Text(350,10)[]{$p_3$} \Text(350,50)[]{$p_4$}
\Text(280,30)[]{$l$} 
\Text(298,-10)[]{$a$}
\end{picture}%
\caption{\sl Feynman graphs for $N~N\to N~N~a$ in the OPE
approximation. On the left, from top to bottom, the direct diagrams $A$, $B$, $C$, $D$. 
On the right, from top to bottom, the exchange diagrams $A'$, $B'$, $C'$, $D'$.}%
\end{floatingfigure}
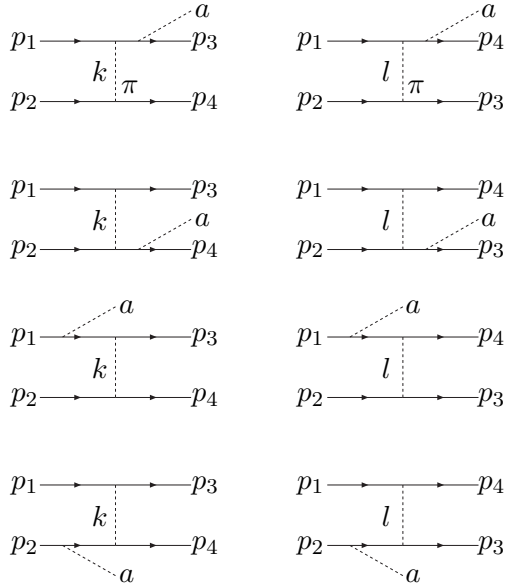

In the OPE approximation, the nucleons interact with each other exchanging one pion $\pi$. 
This interaction is described by the effective vertex $(2mf_{ij}/m_{\pi})\ov N_i \gamma_5 N_j \pi$, 
where $m$ is the nucleon mass ($m_n\simeq m_p$) and 
$f_{ij}\sim 1$ is a phenomenological constant ($i,j=n,p$). 
It depends on whether the pion is chargeless or not, being $f_{np}=\sqrt 2 f_{nn}=-\sqrt 2f_{pp}$, 
as required by the isospin invariance.
Analogously, the axion-nucleon interaction is $(g_{i}/2m)\ov N_i \gamma_\mu\gamma_5 N_i\de^\mu a$,
where the axion-nucleon couplings are defined as $g_{n}=c_n m/f_a$ and $g_{p}=c_p m/f_a$.%
\footnote{In general, the nucleon-pion interaction has the derivative
form $(f_{ij}/m_{\pi})\ov N_i \gamma_\mu\gamma_5 N_j\de^\mu \pi$,
typical of the (pseudo-) Goldstone modes, just as the axion. However,
this interaction can be made pseudoscalar (as in the main text), after
an opportune chiral rotation of the nucleon fields. Yet, this
operation cannot be performed for both the pion and the axion field at
once.  See ref. 17 in \cite{Turn3} for more details}
For the constants $c_{n}\sim c_{p}\sim 1$, they are generally model
independent, since the axion-nucleon interaction arises mainly from
axion-pion mixing.  In the following, we will consider the general
case $g_{n}\neq g_{p}$.

As in the text, we indicate the nucleon momenta with $p_i\simeq(m+\gr
p_i^2/2m, \bf{p_i})$ and that of the axion with $a=(\omega_a,\gr
a)$. Also $\gr k =\gr p_2-\gr p_4$, is the momentum transfer for the
direct diagrams, while $\gr l =\gr p_2-\gr p_3$ refers to the exchange
diagrams.

There are 8 different Feynman graphs that contribute to the process
under examination: 4 direct ($A,B,C,D$) and for 4 exchange
($A^\prime,B^\prime,C^\prime,D^\prime$) diagrams.%
\footnote{The diagrams are the same, and have the same name, as in
reference \cite{Turn3}. We use the same notation here for convenience.}
The total matrix element squared is then given by $\mc
M^2=\left(A+B+C+D+A^\prime+B^\prime+C^\prime+D^\prime\right)^2$.  The
different contributions have the form:
\begin{eqnarray}\label{diagrams1}
	&&X=\frac{1}{|\textbf{k}|^2+m_\pi^2}~\frac{1}{\pm 2 p_i\cdot a+m_a^2}~\frac{2m}{m_\pi^2}~\Omega_X\,,\nn
	&&X^\prime=\frac{1}{|\textbf{l}|^2+m_\pi^2}~\frac{1}{\pm 2 p_i\cdot a+m_a^2}~\frac{2m}{m_\pi^2}~\Omega_{X^\prime}\,,
\end{eqnarray}
where $X$ and $X^\prime$ indicate respectively $A,B,C,D$ and
$A^\prime,B^\prime,C^\prime,D^\prime$, and the index $i$ refers to the
nucleons at which the axion leg is attached.  The "$+$" sign, in the
denominator, applies to the diagrams $A,B,A^\prime,B^\prime$, and the
"$-$" to $C,D,C^\prime,D^\prime$.

For the general nucleon-nucleon axion bremsstrahlung process,
$\alpha~\beta\to\alpha~\beta~a$, the functions
$\Omega_X,\;\Omega_{X^\prime}$ are expressed as:
\begin{eqnarray}\label{diagrams2}
&&\Omega_{A,C}=f_{\alpha\alpha}f_{\beta\beta}~
\ov{u_3}^{(\alpha)}~ \Gamma_{A,C}~ u^{(\alpha)}_1\ov{u_4}^{(\beta)} ~\gamma_5~ u^{(\beta)}_2 g_\alpha 
\\[1ex]\nonumber
&&\Omega_{A^\prime,C^\prime}=f_{\alpha\beta}^2~
\ov{u_4}^{(\beta)}~ \Gamma_{A^\prime,C^\prime}~ u^{(\alpha)}_1\ov{u_3}^{(\alpha)} ~\gamma_5~ u^{(\beta)}_2
\times\left\{
  \begin{array}{ll}
  	g_\beta & ({\rm for}~ A^\prime)\\
  	g_\alpha & ({\rm for}~ C^\prime)
  \end{array}
\right.
\\[1ex]\nonumber
&&\Omega_{B,D,B^\prime,D^\prime}=\Omega_{A,C,A^\prime,C^\prime}(1\leftrightarrow2,3\leftrightarrow4,\alpha\leftrightarrow\beta)\,,
\end{eqnarray}
where the notation $u_i$ is a short for
$u(p_i)$, and the indexes $\alpha$ and $\beta$ stand for the neutron ($n$) or the proton ($p$). 
For example, in the $n~p$ bremsstrahlung the two spinors $u^{(\alpha)},~u^{(\beta)}$ represent
respectively the neutron and proton field and $g_a=g_n,~g_b=g_p$, while for the $n~n$ or $p~p$ bremsstrahlung, 
$u^{(\alpha)}\equiv u^{(\beta)}$ and $g_\alpha=g_\beta$. 
Finally, the matrix functions $\Gamma$ are
\begin{eqnarray}
	&&\Gamma_A=m_a^2+\asl(\psl_3-m)\,,\qquad\qquad \Gamma_C=m_a^2-(\psl_1-m)\asl\,,\nn
	&&\Gamma_{B,D}=\Gamma_{A,C}(1\leftrightarrow 2,~ 3\leftrightarrow 4)\,,
	\qquad \Gamma_{X^\prime}=\Gamma_{X}(3\leftrightarrow 4)\,.
\end{eqnarray}

We have computed the matrix element squared, summed over the nucleon
spin, in the hypotheses (\emph{i}) and (\emph{ii}) of section 3.  As
explained, these imply that the axion three-momentum is negligible
with respect to that typical of the nucleons.

An important consequence of (\emph{i}) and (\emph{ii}) is that $\gr
p_i\cdot \gr a\leq |\gr p_i||\gr a|\ll m \omega_a$, and $m_a^2 \ll m
\omega_a$, so that the second denominator of both equations
(\ref{diagrams1}) is simply $\sim \pm m\omega_a$.  Another consequence
is that the axion mass is always negligible with respect to $|\gr
k|^2$ and $|\gr l|^2\sim 3mT$. This considerably simplifies the
computation of the matrix element squared.

The matrix $\sum_{\rm spin}|\mc M|^2$ contains three different contributions: 
\emph{i}) a term from the product of two direct diagrams, which gives a contribution
proportional to $(|\textbf{k}|^2+m_\pi^2)^{-2}$;
\emph{ii}) a term from the product of two exchange diagrams, which gives a contribution
proportional to $(|\textbf{l}|^2+m_\pi^2)^{-2}$; and finally,
\emph{iii}) a term from the product of one direct and one exchange diagram, which gives a contribution proportional to $(|\textbf{k}|^2+m_\pi^2)^{-1}(|\textbf{l}|^2+m_\pi^2)^{-1}$. 
A straightforward, though very long, calculation leads to:
\begin{equation}\label{result}
	\sum_{\rm spin}|\mc M|^2=
		\frac{32}{9}\frac{m^2}{m_\pi^4}\left(\frac{\gr a^2}{\omega_a^2}\right) \left\{
		\frac{C_{k}|\textbf{k}|^4}{(|\textbf{k}|^2+m_\pi^2)^2}+
		\frac{C_{l}|\textbf{l}|^4}{(|\textbf{l}|^2+m_\pi^2)^2}
		+\frac{C_{kl}|\textbf{k}|^2|\textbf{l}|^2-3\;C_{k\cdot l}|\gr k\cdot\gr l|^2}
			{(|\textbf{k}|^2+m_\pi^2)(|\textbf{l}|^2+m_\pi^2)}\right\},
\end{equation}
where we have averaged over the axion emission angles:
$\br (\hat{\gr k}\cdot \hat{\gr a})^2\kt=\frac{1}{3}$,
$\br (\hat{\gr l}\cdot \hat{\gr a})^2\kt=\frac{1}{3}$.
Observe that the dependence of (\ref{result}) on the axion three-momentum,
satisfies the requirement $|\mc M|^2\to0$ for $\gr a\to 0$, deducible from general considerations on the
(pseudo-)Goldstone nature of the axion, as discussed in section 3.

For the coefficients in (\ref{result}) we found, in general:
\begin{equation}\label{coefficients}
{\arraycolsep=.15em
	\begin{array}[b]{rclrcl}     
	C_{k}&=&12 f^4 (g_\alpha^2+g_\beta^2)\,, & C_{l}&=&3f_{\alpha\beta}^4(3 g_\alpha^2+3g_\beta^2+2g_\alpha g_\beta)\,,\\[1ex]
	C_{kl}&=&12 f^2 f_{\alpha\beta}^2(g_\alpha^2+g_\beta^2)\,,\qquad & C_{k\cdot l}&=&8f^2 f_{\alpha\beta}^2(g_\alpha^2+g_\beta^2+g_\alpha g_\beta)\,, 
	\end{array}
}
\end{equation}
where $f=f_{nn}=-f_{pp}$.%
\footnote{Observe that there is a relative minus sign between direct
and exchange diagrams for the case of $n~n$ or $p~p$ process. This is
taken into account in the definition of the parameters
(\ref{coefficients}). In fact, in $C_{kl}$ and $C_{k\cdot l}$,
$f_{\alpha\alpha}f_{\beta\beta}$ should appear in place of $f^2$. So,
for the $n~p$ process, $C_{kl}$ and $C_{k\cdot l}$ are written in
(\ref{coefficients}) with the wrong sign. However, this is compensated
by the sign of $C_{kl}$ and $C_{k\cdot l}$ in (\ref{result}), which
should be the opposite for the $n~p$ bremsstrahlung.}

In the simple cases of $n~n$ or $p~p$ processes, relations
(\ref{coefficients}) lead to the simple result:
\begin{equation}\label{np}
	C_{k}=C_{l}=C_{kl}=C_{k\cdot l}=24f^4g_n^2\,,
\end{equation}
that agrees with eq (\ref{matrix0}) (observe that in this case
$S=1/4$, and remember that $\alpha_\pi=(2fm/m_\pi)^2/4\pi$).  The more
complicated $n~p$ scattering, instead, requires:
\begin{equation}
{\arraycolsep=.15em
	\begin{array}[b]{rclrcl}     
	C_{k}&=&12 f^4 (g_n^2+g_p^2)\,,\qquad & 
	C_{l}&=&12f^4(3 g_n^2+3g_p^2+2g_n g_p)\,,\\[1ex]
	C_{kl}&=&24 f^4(g_n^2+g_p^2)\,, & 
	C_{k\cdot l}&=&16f^4(g_n^2+g_p^2+g_n g_p)\,,
	\end{array}
}
\end{equation}
which leads to the matrix element squared:
\begin{eqnarray}\label{MatrNP}
	\sum_{\rm spin}|\mc M|^2&=&
		\frac{128}{3}\frac{m^2f^4}{m_\pi^4}\left(\frac{\gr a^2}{\omega_a^2}\right) \left\{
		\frac{ (g_n^2+g_p^2)|\textbf{k}|^4}{(|\textbf{k}|^2+m_\pi^2)^2}+
		\frac{(3 g_n^2+3g_p^2+2g_n g_p)|\textbf{l}|^4}{(|\textbf{l}|^2+m_\pi^2)^2}\right.\nn
		&&{}\qquad\qquad\qquad\quad+\left.\frac{2 (g_n^2+g_p^2)|\textbf{k}|^2|\textbf{l}|^2-\frac43(g_n^2+g_p^2+g_n g_p)|\gr k\cdot\gr l|^2}
			{(|\textbf{k}|^2+m_\pi^2)(|\textbf{l}|^2+m_\pi^2)}\right\}
\end{eqnarray}
In the limit $m_a=0$ this corresponds to the result in
ref. \cite{Turn3}.

\bigskip

In what follows, we will refer to the $n~n$ process (the $p~p$ process
is equivalent), unless we specify otherwise.

For the computation of the axion emission rate, it is rather
convenient to introduce the new set of variables:%
\footnote{The notation here strictly follows the appendix of ref. \cite{RS}.}
the center of mass momenta $\gr p_{1,2}=\gr P\pm\gr p_i$, $\gr
p_{3,4}=\gr P^\prime\pm\gr p_f$, where $\gr P=\frac12(\gr p_1+\gr
p_2)$ and $\gr P^\prime=\frac12(\gr p_3+\gr p_4)$; the cosine of the
nucleon scattering angle $z=\gr{\hat{p}}_i\cdot\gr{\hat{p}}_f$; the
adimensional parameters: $u=\gr p_i^2/mT$, $v=\gr p_f^2/mT$,
$y=m_\pi^2/mT$, $x=\omega/T$, and $q=m_a/T$.  Thus the matrix element
squared can be conveniently expressed as:
\begin{equation}\label{M2}
	\{\mc{M}^2\}_{nn}=\zeta^2~\frac{64\pi^2\alpha_\pi^2}{3 m^2}~g_{n}^2 ~\eta\,,
\end{equation}
where $\zeta=(1-q^2/x^2)^{1/2}$ is the axion velocity, and
$\eta=(\eta_{\gr k}+\eta_{\gr l}+\eta_{\gr{kl}}-3\eta_{\gr{k\cdot
l}})$,
\begin{eqnarray}\label{eta}
	&&\eta_{\gr k}=\left(\frac{u+v-2z\sqrt{uv}}{u+v-2z\sqrt{uv}+y}\right)^2,
	\qquad  \eta_{\gr l}=\left(\frac{u+v+2z\sqrt{uv}}{u+v+2z\sqrt{uv}+y}\right)^2\,,\nn
	&&\eta_{\gr{kl}}=\frac{(u+v)^2-4uvz}{(u+v+y)^2-4uvz}~,\qquad\quad \eta_{\gr{k\cdot l}}=
        \frac{(u-v)^2}{(u+v+y)^2-4uvz}\,. 
\end{eqnarray}
Observe that the axion velocity $\zeta$ measures the only contribution
of finite axion mass to the matrix element squared.

For negligible axion three-momentum (with respect to that of the
nucleons) the delta function in (\ref{dN}) is simply $\delta^3(\gr
P-\gr P^\prime)\delta(u-v-x)/T$.  Moreover the axion distribution is
isotropic, since we have already averaged over the axion momentum
directions. Hence, $d\Pi_a=(T/2\pi)^2\zeta x\,dx$, and expression
(\ref{dN}) can be recasted in the more convenient form:
\begin{equation}\label{N}
	T\;\frac{d\mc N}{dx}=\frac{35}{128}\tilde Q~
	\zeta^{3}x\int du ~\int dv \sqrt{uv}e^{-u}\delta(u-v-x)~
	\frac12\int_{-1}^1\eta~dz\,,
\end{equation}
where the constant factor
\begin{equation}
	\tilde Q=\frac{32}{105}\frac{\rho^2 T^{7/2}\alpha_\pi^2}{\pi^{3/2}m^{13/2}}~g_n^2
	\simeq 3.4 \times 10^{42} T_{\mev}^{7/2}\;\rho_{12}^2\;g_n^2~{\rm erg~ cm^{-3}~s^{-1}}
\end{equation}
is related to the axion energy emission rate for $m_a=m_\pi=0$
(\ref{Q0}), as we are going to show.  In fact we can write the axion
energy emission rate as:
\begin{equation}\label{Q}
	Q=T\int_q^\infty x\frac{d\mc N}{dx} dx\,.
\end{equation}
Observe that this corresponds to relation (B-6) in ref.\cite{RS},
except for the correction factor $\zeta^{3}$ in~(\ref{N}) and the
lower integration limit $q$.  In the limit of negligible pion mass,
$\eta$ reduces to $\eta_0=3(1-\eta_{\gr{k\cdot l}})$. We see that, for
$m_a=0$, the first term in $\eta_0$ contributes to $Q$ as $3\tilde Q$.
If we define $Q_{\gr k\cdot\gr l}$ as the contribution to $Q$ from
$3\eta_{\gr k\cdot\gr l}$, then, (still in the limit $m_a=0$), the
second term in $\eta_0$ contributes to $Q$ as $-\beta \tilde Q$, where
$\beta=Q_{\gr k\cdot\gr l}/\;\tilde Q\simeq1.31$. The same argument
can be repeated for $\mc N$, defining $3\tilde \mc N$ as the
contribution to $\mc N$ from the first term in $\eta_0$, $\mc N_{\gr
k\cdot\gr l}$ as the contribution form the second term in $\eta_0$,
and $\beta^\prime=\mc N_{\gr k\cdot\gr l}/\;\tilde \mc N\simeq1.02$.
Finally, we recover (\ref{Q0}) and (\ref{N0}) in the form:%
\footnote{It is useful to notice that $\int\!
du\,dv\,dx\,x^n\,\sqrt{uv}\,e^{-u}\delta(u-v-x)=8/5$ for $n=1$, and
$128/35$ for $n=2$.  Thus $\tilde \mc N=(7/16)~ \tilde Q/T.$}
\begin{equation}\label{QN0}
	Q_0=\tilde Q(3-\beta)\,, \qquad \mc N_0=\tilde{\mc N}(3-\beta^\prime)\,.
\end{equation}

Notice that the direct substitution of $|\hat{\gr k}\cdot\hat{\gr
l}|^2$ with $\beta/3$ in the matrix (\ref{result}) (see,
e.g.,\cite{Turn3}) can be incorrect, even in the limit $m_a=0,~y\ll
1$, and is strictly valid only concerning the contribution to the
axion energy emission rate.  For example, the substitution above would
have brought the result $\mc N_0=\tilde{\mc N}(3-\beta)\simeq
1.69\tilde{\mc N}$, instead of $\mc N_0=\tilde{\mc
N}(3-\beta^\prime)\simeq 1.98\tilde{\mc N}$.  However the error that
results is less than $20\%$, which is usually negligible with respect
to other approximations necessary for the calculation (see, e.g., the
discussion in~\cite{raffBook}, page 120).

\bigskip

For the sake of comparison with other papers, we consider the
dynamical contribution to the axion energy emission rate $Q$, in the
limit of zero pion mass, in the $n~p$ bremsstrahlung.  Suppose
$g_n=g_p=g_N$. Then, from (\ref{MatrNP}), and substituting $|\hat{\gr
k}\cdot\hat{\gr l}|^2=\beta/3$, we find
\begin{equation}
	\{\mc M^2\}_{np}=\zeta^2\frac{256}{3}\frac{m^2f^4}{m_\pi^4}(7-2\beta)\,,
\end{equation}
that means $\{\mc M^2\}_{np}=4[(7-2\beta)/(3-\beta)]\{\mc M^2\}_{nn}$,
about $10.4$ times larger. This agrees with the results
in~\cite{Turn3,Turn1}, in the limit of zero axion mass.  The analogous
contribution to the axion emission rate $\mc N$, has the same
expression as above, but with $\beta\to \beta^\prime$, and leads to
$\mc N\sim 10$ times larger.  The average energy of the emitted axions
in the $n~p$ bremsstrahlung, is then slightly larger, $\ov
\omega_a\sim 2.02$, with respect to the $n~n$ or $p~p$ process.

\bigskip

We finally consider the mean free path.  In this case the relevant
process is the axion absorption by the nuclear medium $N~N~a\to N~N$.
Thus the axion energy appears in the $\delta$ function with the
opposite sign with respect to (\ref{N}).  The axion mean free path can
then be written as
\begin{equation}
  \lambda^{-1}=\frac{1}{2\,\zeta\,x\, T}\frac{d\mc N(-x)}{d\Pi_a}=
  \frac{2\pi^2}{\zeta^2x^2\, T^3}\frac{d\mc N(-x)}{dx}\,.
\end{equation}
Eliminating the integration over $v$, by virtue of the $\delta$
function, we get
\begin{equation}
	 \lambda^{-1}=\frac{35~\pi^2}{64}\frac{\tilde Q}{T^4}\zeta~x^{-1}~f(x,y)\,,
\end{equation}
where
\begin{equation}\label{f}
	f(x,y)=\int_0^\infty du \left((u+x)u\right)^{1/2} e^{-u}~\frac12\int_{-1}^1\eta~dz\,.
\end{equation}
Observe that, if $\eta$ is constant, $f(x,y)$ can be analytically
expressed in terms of the modified Bessel functions.  In
ref.\cite{Turn1}, it is assumed $\eta\simeq 3-\beta$, which is a
pretty good approximation in the limit $y\ll 1$ (see the discussion
above).  In this case, expression (\ref{f}) reduces to
$$
	\frac12(3-\beta)\;x\; e^{x/2} \;K_1(x/2)\,,
$$
and, consequently, 
\begin{equation}
	 \lambda^{-1}\simeq\frac{35~\pi^2}{128}\frac{Q_0}{T^4}\zeta~e^{x/2} K_1(x/2)=
	 2.5\times 10^6~T_{\mev}^{-1/2}\;\rho_{12}^2\;g_n^2\;\zeta~e^{x/2} 
	 K_1(x/2)~{\rm cm^{-1}}\,,
\end{equation}
where we have used $\tilde Q(3-\beta)=Q_0$.  This expression leads
directly to the results (\ref{lambda0}) and (\ref{lma}) in section 3.

\end{document}